# 2D MoS$_2$-Graphene-based multilayer van der Waals heterostructures: Enhanced charge transfer and optical absorption, and electric-field tunable Dirac point and band gap

Liang Xu[1,2], Wei-Qing Huang[1*], Wangyu Hu[2], Bing-Xin Zhou[1], Anlian Pan[1], Gui-Fang Huang[1#]

[1] *Department of Applied Physics, School of Physics and Electronics, Hunan University, Changsha 410082, China*

[2] *School of Materials Science and Engineering, Hunan University, Changsha 410082, China*

**Abstract**: Multilayer van der Waals (vdWs) heterostructures assembled by diverse atomically thin layers have demonstrated a wide range of fascinating phenomena and novel applications. Understanding the interlayer coupling and its correlation effect is paramount for designing novel vdWs heterostructures with desirable physical properties. Using a detailed theoretical study of 2D MoS$_2$-graphene (GR)-based heterostructures based on state-of-the-art hybrid density functional theory, we reveal that for 2D few-layer heterostructures, vdWs forces between neighboring layers depend on the number of layers. Compared to that in bilayer, the interlayer coupling in trilayer vdW heterostructures can significantly be enhanced by stacking the third layer, directly supported by short interlayer separations and more interfacial charge transfer. The trilayer shows strong light absorption over a wide range (<700 nm), making it very potential for solar energy harvesting and conversion. Moreover, the Dirac point of GR and band gaps of each layer and trilayer can be readily tuned by external electric field, verifying multilayer vdWs heterostructures with unqiue optoelectronic properties found by experiments. These results suggest that tuning the vdWs interaction, as a new design parameter, would be an effective strategy for devising particular 2D multilayer vdWs heterostructures to meet the demands in various applications.

[*]. Corresponding author. *E-mail address:* wqhuang@hnu.edu.cn
[#]. Corresponding author. *E-mail address:* gfhuang@hnu.edu.cn

Vertically stacked two-dimensional (2D) heterostructures made by layer-by-layer (such as graphene (GR) and related 2D materials) in a precisely chosen sequence, has led to promising new materials for various fields, including nanoelectronics, energy conversion and storage, nanooptics, and catalysis [1-3]. These heterogeneous stacks have unusual properties that are not present in individual layers and encompass a wide spectrum of physical and chemical phenomena exemplified by new van Hove singularities[4-7], Fermi velocity renormalization[8, 9], unconventional quantum Hall effects[10], Hofstadter's butterfly pattern [11-14], and others. Peculiar electronic[15-17] and optoelectronic properties[18, 19] have been revealed in diverse 2D heterostructures based on layered materials such as GR, semiconducting transition metal dichalcogenides (TMDs) and insulating hexagonal boron nitride (h-BN). Interestingly, these heterostructures are also tunable hyperbolic metamaterial[20] and memory device with ultrahigh on/off ratio[21], or have gate-induced superconductivity[22] and high-temperature superfluidity[23].

In recent five years, strong interlayer coupling and its effects on novel physical phenomena and diverse potential applications of vdWs heterostructures built from 2D materials has become a hot topic in various disciplines [2, 15, 16, 18, 19, 24-28]. The multilayer vdW heterostructures can be experimentally prepared by using state-of-the-art atomically thin 2D materials synthesis techiques, such as, chemical vapour deposition (CVD) to directly grow through the vertical layer-by-layer growth mode[29, 30], and exfoliation techniques with roll-to-roll assembly[2] and co-segregation method[31]. Interlayer coupling in 2D heterostructures has proved to be van der Waals (vdW) interaction, which becomes a promising approach for modifying the materials properties of 2D heterostructures while maintaining many of the desired properties of pristine individual layers, such as GR's high charge carrier mobility[1, 2]. Although vdW interactions in these 2D heterostructures are assumed to be responsible for the emergence of various novel properties and new phenomena, it is still a vague generality of the interactions between adjacent layers. Advancing 2D heterostructures of knowledge and related technologies will require a detailed understanding of interlayer interactions. Despite contributions from several disciplines and many groups, quantitative or even qualitative accounting of the interactions in 2D multilayer heterostructures is rarely achieved.

Herein we aim at elucidating the effects of interlayer interactions on the properties of 2D

multilayer vdW heterostructures. GR-MoS$_2$-based heterostructures are chosen as prototype 2D multilayer vdW heterostructures because that they have showed enormous perspectives in technological applications, and are the most experimentally studied 2D multilayer systems[1, 2]. Moreover, GR and MoS$_2$ are likely to be the most common component in future vdW heterostructures and devices due to their stablility at ambient conditions, making them useful for passivation of other unstable 2D materials, such as ultrathin black phosphorus [32]. Indeed, GR is usually used to be the outermost layer in 2D multilayer heterostructures owing to its the highest mechanical strength and crystal and electronic quality among the 2D crystals. Similarly, 2D MoS$_2$ monolayer have an optical bandgap in the near-infrared to visible spectral range and exhibit extremely strong light-matter interactions, making it particularly exciting for novel optoelectronic and photovoltaic applications[2]. We present a detailed theoretical study of the properties of several GR-MoS$_2$-based 2D multilayer heterostructures, based on state-of-the-art density functional theory (DFT) calculations, with explicit inclusion of vdW effects. It is demonstrated for the first time that compared to that in bilayer, the interlayer coupling between neighboring layers in trilayer vdW heterostructures can be obviously enhanced by the addation of third layer, which is directly supported by short interlayer separations and stronger charge transfer. Due to its unique near-gap electronic structure caused by vdW interactions, the trilayer displays strong light absorption over a wide range (<700 nm), making it very potential for highly active photodetectors, solar energy harvesting and conversion. Moreover, perpendicular electric field can induce tunable Dirac point of GR and band gaps of each layer and trilayer, indicating realizing the band engineering at interface in 2D multilayer vdW heterostructures. These findings may open a way to tune the interlayer coupling and properties of few-layer vdWs heterostructures by change the number of layers, besides choosing appropriate 2D layers.

**Computational Methods**. All our calculations are based on the DFT in conjunction with the projector-augmented-wave (PAW) potential[33] as implemented by the Vienna ab initio simulation package (VASP)[34]. The Perdew-Burke-Ernzerhof exchange-correlation functionals with the approach of Grimme (DFT-D3) is adopted to correct the weak van der Waals[35]. The screened hybrid Heyd-Scuseria-Ernzerhof 2006 (HSE06) functional containing a mixture of the exact exchange (17.5%) has been employed to get accurate electronic structures and optical properties.

The cutoff energy for the plane-wave basis set is 500 eV and the first Brillouin zone is sampled with a Monkhorst-Pack grid of $15 \times 15 \times 1$. All atomic positions are fully relaxed until the force is less than 0.01 eV/Å. The criterion for the total energy is set as $1\times10^{-6}$ eV. A vacuum space of 15 Å along the z direction is used to decouple possible periodic interactions.

**Interfacial charge transfer in 2D multilayer heterostructures**. Although the interaction between vdW stacked 2D crystals is relatively weak, electron orbitals still extend out of the plane, even overlap with those of an adjacent 2D crystal, such as electrical coupling in the $WSe_2/MoS_2$ hetero-bilayer[36]. In the case of monolayer ZnO on $MoS_2$, the interfacial charge redistributions in six stacking patterns with high symmetries can be divided into three types (see Fig. S1 for details), which depend on the O (Zn) -S atoms relative positions and interlayer spacing. Among them, the A-B stacking $ZnO/MoS_2$ hetero-bilayer with Zn atoms positioned over S atoms displays the most obvious charge redistributions at interface (Fig. S1 (a2) and Fig. 1(e)) due to its smallest interlayer spacing and relatively big electronegativity difference between Zn and S atoms (Table S1), and hereafter this stacking pattern is chosen as a representative to construct multilayer heterostructures.

To shed light on the electron transfer at each interface of the 2D vdW multilayer heterostructures, GR is put on top of $ZnO/MoS_2$ hetero-bilayer to form a hetero-trilayer, as is shown in Figs. 1 (a) and (c), which has been usually done in experiments [37]. This enables us to study the influence of added layer on the electron transfer at initial interfaces of multilayer heterostructures. The amounts of electron transfer can be visualized in a very intuitive way, by the three-dimensional charge density difference $\Delta\rho= \rho_{trilayer} / \rho_{bilayer} - \rho_{monolayer}$, where $\rho_{trilayer}/\rho_{bilayer}$ and $\rho_{monolayer}$ are the charge densities of hetero-trilayer/hetero-bilayer, and free-standing monolayers (i.e., monolayer GR, ZnO, and $MoS_2$) in the same configuration, respectively. Compared to that in hetero-bilayer (Fig. 1 (e)), the interfacial charge changes at the interface of ZnO and $MoS_2$ in hetero-trilayer are greatly reinforced, even the Mo atoms sandwiched between two S atom layers gain some electrons from S atoms at top layer, as is shown in Fig. 1 (c). The reinforcement of interfacial electron transfer is only due to the addition of GR on the ZnO layer, indicating that the interaction between ZnO and $MoS_2$ is strengthened by GR, in agreement with the reduction of their distance from 2.890 Å in hetero-bilayer to 2.835 Å in hetero-trilayer (See Table 1). In the meantime, GR/ZnO/MoS$_2$ hetero-trilayer can be viewed as monolayer MoS$_2$ grown on GR/ZnO hetero-bilayer. Similarly, the charge transfer at the GR-ZnO interface is also obviously enhanced (see Figs. 1 (b) and (c)), and their distance is decreased (3.134 vs. 3.153 Å, Table 1), owing to monolayer MoS$_2$ stacking. To offer more details of charge redistribution, Fig. 1 (d) plot the planar averaged charge density difference along the direction perpendicular to GR plane. This further confirms that in the multilayers, the amount of charge transfer between two neighboring layers is much large than in the corresponding hetero-bilayer.

A further charge analysis based on the Bader method can give the quantitative results of charge transfer in the 2D vdW multilayer heterostructures, which are listed in Table 1. In the two hetero-bilayer systems, 0.027 electrons transfer from GR sheet to ZnO layer, and 0.066 electrons transfer from ZnO layer to MoS$_2$ layer. The spontaneous interfacial charge transfer in hetero-bilayer can be simply rationalized in terms of the differences of the work functions between different layers, and interfacial distance. The larger difference in work functions and the smaller interfacial distance, the more charge transfer[38]. The work function difference (0.27 eV) between GR and ZnO monolayer is smaller than that (0.74 eV) between ZnO and MoS$_2$ layer (See Table 2); thus, the transfer amount of charge in the former is much less than in the latter. In fact, interface charge doping of GR as a consequence of work function difference between GR and the substrate are well documented and well understood especially for metals [39, 40]. Unexpectedly, the amounts of charge transferred of GR, ZnO, and MoS$_2$ layers in hetero-trilayer are 0.064, 0.197, and 0.261 electrons, respectively, much larger than those in hetero-bilayer. The larger amounts of charge transferred in the hetero-bilayer further corroborates that the interfacial interaction between the neighboring layers can be strengthened by adding other 2D layers.

To compare the relative stability of these 2D vdW heterostructures, their interface binding energies, $\Delta E_f$, are estimated according to the following equation, $\Delta E_f = E_{trilayer}/E_{bilayer} - E_{monolayer}$, where $E_{trilayer}/E_{bilayer}$ and $E_{monolayer}$ represent the total energy of hetero-trilayer/hetero-bilayer, and free-standing monolayers (i.e., monolayer GR, ZnO, and MoS$_2$), respectively. By this definition, the negative $\Delta E_f$ indicates that 2D vdW heterostructures is stable. The lower $\Delta E_f$ will make the heterostructures to approach to the lower energy state and therefore be steadier, sugggesting that the heterostructures can be experimentally synthesized more easily.

Table 1 shows that the interface binding energy of GR/ZnO/MoS$_2$ hetero-trilayer is -5.853 eV, smaller than those of GR/ZnO and ZnO/MoS$_2$ hetero-bilayers, demonstrating that trilayer heterostructure is more stable than bilayer one. This is consistent with its smaller interface distance and larger amount of charge transferred.

**Electronic structures in 2D multilayer heterostructures.** Interfacial charge transfer in 2D multilayer heterostructures due to vdW interactions is bound to affect the electronic structure of each layer, thus leading to the systems having unusual properties and new phenomena, which are beyond the individual layers. This can be confirmed by comparing the band structures of monolayer, bilayer, and trilayer. Figs. 2 (a-c) display the projected band structures of GR/ZnO, ZnO/MoS$_2$ and GR/ZnO/MoS$_2$ vdW heterostructures. Pure GR is a zero-gap semiconductor, ZnO and MoS$_2$ monolayers have the band gap of 3.15 eV and 1.97 eV, respectively (Fig. S2). To rule out the influence caused by the lattice aberrance to overcome the mismatch in supercell, the band structures of three monolayers with the same parameters as in trilayer vdW heterostructure are calculated (Fig. S2 (a1-c1)). Obviously, the band gaps and near-gap electronic structures of bilayer and trilayer vdW heterostructures have been tuned by the weak vdW interactions between adjacent layers, although the shapes of band structures of all the three layers are well-preserved (Figs. 2 and S2). As is expected, GR's band gap of 5.7 meV is opened in GR/ZnO bilayer (Fig. 2(a)). At the Fermi level, there is no states from ZnO layer, and consequently it is governed by the GR states at the Dirac-point. The ZnO/MoS$_2$ bilayer is an indirect gap semiconductor hybrid (E$_g$≈1.2 eV), in which the conduction band minimum (CBM, labeled by L$^b$, is only composed of Mo 4d states, as shown in Figs. 2(e) and S3) appears at the Γ point, whereas the valence band maximum (VBM, labeled by H$^b$) at the one point within the KΓ line (Fig. 2(b)). The VBM is the hybridization of Mo 4d, S 3p and O 2p states, in which the contribution of S atom (S$_1$ atom) facing ZnO layer is bigger than that away (S$_2$ atom) (see Figs. 2(d) and S3). This leads to the upshift of VBM, thus a smaller band gap of ZnO/MoS$_2$ bilayer compared to MoS$_2$ monolayer.

Fig. 2 (c) shows that the near-gap electronic structure of GR/ZnO/MoS$_2$ vdW trilayer heterostructure changes dramatically due to large amount of charge transferred between the adjacent layers. Firstly, the Fermi level of trilayer moves up to CBM of MoS$_2$ layer, from VBM of pure monolayers and the middle gap of ZnO/MoS$_2$ bilayer. This is due to more electrons transferred to $MoS_2$ layer from GR and ZnO monolayer, as displayed in Fig. 1(c). Secondly, the Dirac point of GR shifts over the Fermi level, thereby indicating p-type doping of GR, which is in consistent with the fact that GR losses more electrons in trilayer than in pure $ZnO/MoS_2$ bilayer (Fig. 1 and Table 1). Thirdly, the energy gap between $H^t$ and $L^t$ is decreased from 1.2 to 0.97 eV. In addition, the band gap of GR in trilayer is also reduced in contrast to in GR/ZnO bilayer. The decrease of the two band gaps is ascribed only to the variation of interfacial interactions due to the third layer addition. Careful observations of Figs. 2 (d) and (e) shows that compared to that in bilayer, the electronic density distribution of CBM of $ZnO/MoS_2$ in trilayer is nearly unchanged, whereras the electronic density distribution of VBM of $ZnO/MoS_2$ in trilayer is obviously altered by the stacking GR sheet: the ZnO's weight in electronic density of VBM of $ZnO/MoS_2$ in trilayer is much smaller than in bilayer. These results demonstrate that the near-gap electronic properties of 2D vdW heterostructure can be tuned by changing the number of hetero (homo)-layers, which is significance to design 2D heterostructures with tunable optoelectronic properties [41, 42]. In fact, a recent experiment proved that the electronic structure in few-layer $MoSe_2$ nanostructures are highly dependent on the number of layers[43].

**Electrostatic potential in 2D multilayer heterostructures**. Additional insights into the effect of layer number in 2D vdW heterostructure are obtained by analyzing the electrostatic potential distribution. The planar averaged self-consistent electrostatic potential for the bilayer and trilayer 2D vdW heterostructures as a function of position in the z-direction is shown in Fig. 3 (Upper panel). The electrostatic potential at the plane of GR sheet is remained about the same regardless of layer number in 2D GR-based heterostructure, while that at the plane of ZnO sheet in trilayer heterostructure is slightly elevated compared with in bilayer one. In contrast, the electrostatic potentials at the planes of two S atom layers are obviously lowered by adding the third layer (i.e., GR sheet). The variation of the electrostatic potential at different layers is caused by the electron transfer, and the change value depends on the amount of charge transferred (see Table 1). As a results, the potential difference between S atom layer and ZnO sheet is apparently increased from 6.71 eV in bilayer to 9.52 eV in trilayer heterostructure. The larger potential step that act as transport barriers between adjacent layers is expected to significantly influence charge transport and optical property of 2D multilayer vdW heterostructures. For this specific vertical trilayer devices, monolayer $ZnO_2$ is used as tunnel barriers with GR and $MoS_2$ sheets serving as both electrodes——— to form a electronic device: field effect tunnelling transistors. Lower panel in Fig. 3 clearly displays two stripes with low potential at the GR and $MoS_2$ layers, while the alternation of positive and negative local potential along ZnO layer. Therefore, the electron or hole will migrate in different layers in a different way, demonstrating that the properties of 2D multilayer vdW heterostructures can be designed through controlling the electronic potential distribution by choosing appropriate 2D materials, the stacking sequence or layer number. As an example, the optical property modulation of 2D multilayer vdW heterostructures will be discussed next.

**Optical properties of 2D multilayer heterostructures**. In the 2D multilayer heterostructures, the vdWs forces will attract adjacent layers, thus tuning their near-gap electronic structure and photoelectric properties [41, 42]. We first investigate the optical properties. Figure 4 shows the absorption spectra of monolayer, bilayer and trilayer vdW heterostructures calculated by the Fermi golden rule within the dipole approximation. Due to its large band gap (~3.15 eV), ZnO monolayer can only absorb ultraviolet (UV) light. $MoS_2$ monolayer can absorb a small part of the visible (vis-) light (< 500 nm) corresponding to its small band gap, and the absorption intensity is quite low. This drawback can be improved to a certain extent as the two layers form a vertical $ZnO/MoS_2$ bilayer (red curve in Figure 4). Although the absorption spectrum of GR covers the entire UV to far-infrared range, the responsivity and intensity of GR are very low due to the small optical absorption of a monolayer of carbon atoms. As for GR/ZnO bilayer, its absorption curve of is almost the same as that of GR monolayer because the electronic structure near the Fermi level of the former is only composed of the latter. Surprisingly, trilayer's absorption spectrum is profoundly changed due to the addition of third layer. Figure 4 shows that the $GR/ZnO/MoS_2$ trilayer is an excellent photon absorber over a very broad photo-irradiation range including UV, vis-light, and near infrared radiation. In particular, the light absorption intensity of trilayer is dramatically enhanced over a wide range (<700 nm). The enhanced light absorption of trilayer can be attributed to its unique near-gap electronic structure caused by vdW interactions between adjacent layers. The results imply that the trilayer vdW heterostructures have a high utilization efficiency of solar energy, demonstrating that they would be highly active photodetectors or photocatalyst under visible light [3]. More recently, near-unity absorption in vdWs semiconductors has been observed experimentaly[44], in agreement with the theoretical results given here.

**Electronic properties of 2D multilayer heterostructures**. A great deal of studies have demonstrated that 2D vdW heterostructures have novel electronic and others properties, thus multiple device functionalities [2, 45]. In the GR/ZnO/MoS$_2$ trilayer, the addition of third layer give rise to not only p-type doping of GR but also an upshift of the Fermi level at the CBM of MoS$_2$ (Fig. 2(c)). Without external electronic field, the Fermi energy of trilayer is 0.95 eV, much higher than those of monolayers and bilayers which are less than zero, as given in Fig. 5 (a). The increase of Fermi energy is very important for designing the electronic devices with high performance. For example, an increase of Fermi energy of field-effect tunneling transistor effectively lowers the tunnel barrier, even if no bias is applied [46]. An ingenious experiment has shown that direct grown multijunction heterostructures based on GR, MoS$_2$, MoSe$_2$ and WSe$_2$ exhibit resonant tunnelling of charge carriers, which leads to sharp negative differential resistance (NDR) at room temperature[15]. To explore the influence of the external field on the electronic properties of 2D multilayer heterostructures, the projected band structures of GR/ZnO/MoS2 trilayer under various vertical external field are calculated. We find that the band structure of GR/ZnO/MoS2 trilayer experiences a complicated change as the direction and magnitude of external electric field are altered (Figs. S5 and S6).

To gain further insight, Figs. 5 (b-d) show the Fermi energy variation, shift of Dirac point, and evolution of the band gap in the GR/ZnO/MoS$_2$ trilayer in the presence of an applied external perpendicular electric field ($E_\perp$). Here, the direction of electric field $E_\perp$ from GR to MoS$_2$ layer is taken as the forward direction. In the case of a reverse $E_\perp (< 0)$, the Fermi energy of the GR/ZnO/MoS$_2$ trilayer nonlinearly decreases from 0.95 to -0.17 eV with increasing $E_\perp (< -1.0$ V/Å) (Fig. 5 (b)). In contrast, by applying a forward $E_\perp (> 0)$, its Fermi energy decreases firstly ($E_\perp < 0.10$ V/Å), and then increases (except $E_\perp = 0.80$ V/Å) with the $E_\perp$ increase. The tunable Fermi energy of trilayer by the external electric-field is of very important to the photoelectric devices. Interestingly, Fig. 5(c) shows that with respect to the Fermi level, the Dirac point of GR is monotonically reduced from 5.1 to -3.3 eV as the $E_\perp$ changes from -1.0 to 0.7 V/Å. Note that the characteristic Dirac point at K is destroyed when the $E_\perp > 0.7$ V/Å. In a freestanding GR sheet, the Dirac point coincides with the Fermi level, interfacial interaction generally staggers them. The shift upwards (downwards) of Dirac point with respect to the Fermi level means that holes (electrons) are donated by the adjacent layer (here is ZnO monolayer) to GR sheet which becomes p-type (n-type) doped. The tunable of Dirac point is very useful because it is central to various phenomena in condensed-matter physics, from massless electrons in GR to the appearance of conducting edge states in topological insulators. The charge transfer between different layers is also reflected in the shift of $L^t$ point (CBM of ZnO/MoS$_2$ in trilayer) with respect to Fermi level: the energy of $L^t$ point increases monotonically from -1.9 to 2.8 eV as the $E_\perp$ alters from -1.0 to 1.0 V/Å (Fig. 5(c)). The variation of Dirac and $L^t$ points in the trilayer is due to the combined effect of both the external electric-field and vdW interactions between layers.

Opening and tailoring the band gap of GR has long been a hot topic in the GR research. In the GR/ZnO/MoS$_2$ trilayer, the band gap of GR is opened about 4.8 meV due to interfacial interaction. Fig. 5(d) shows that under the external electric-field, the band gaps of GR and MoS$_2$ layers are nonlinearly changed, and their variation is not symmetrical under different direction of the external electric-field because of the inherent asymmetry of the trilayer. Specifically, GR's band gap decreases from 7.3 to 0.1 meV as $E_\perp$ changes from -1.0 to 0.4 V/Å. Unexpectedly, when $E_\perp$=0.5 V/Å, GR's band gap is sharply enlarged to 9.4 meV, and then reduced to very small value (0.09 meV) with increasing $E_\perp$ (< 0.7 V/Å). The reason for the abrupt change of GR's band gap needs more discussions. For the MoS$_2$ layer in trilayer, by contrast, its band gap variation with external electric-field is quite different: the band gap of MoS$_2$ layer decreases as an external electric-field is applied to trilayer, even regardless of its direction. In particular, the band gap of MoS$_2$ is obviously decreased from 0.97 eV to zero when $E_\perp$ alters from 0 to -0.5 V/Å, whereas it only reduces slightly as $E_\perp$ increases from 0 to 0.6 V/Å, and then rapidly decreases from 0.95 eV to 0.5 eV with further increasing $E_\perp$. The external electric-field effect on the band gap change of different layers in trilayer is essentially different from those in MoS$_2$-based bilayer, such as 2D black phosphorus (BP)/ MoS$_2$ heterostructure, in which the external electric field exerts little influence on the respective band gap of BP and MoS$_2$ [47]. Overall, different regulatory effects on the properties of different layers under the external electric field make the 2D vdWs multilayers having huge potential in various applications, such as photocatalysis and novel photoelectric devices.

**Discussions and Summary**. Depending on the particular structure we reveal that typical 2D vdWs GR/ZnO/MoS$_2$ trilayer renders enhanced charge transfer and optical absorption, as well as electric-field tunable Dirac point and band gap. To rule out the effect of specific layer sequences, we change the arrange order of GR and ZnO layers in trilayer. Figure S6 clearly renders that in the ZnO/GR/MoS$_2$ trilayer, the amount of charge transfered between adjacent layers is much more prominent than their corresponding bilayers, especially for that of ZnO layer (Figure S6(d)). Accordingly, the separation of the layers in trilayer is smaller than those in bilyer. Moreover, the optical absorption of ZnO/GR/MoS$_2$ trilayer is significantly increased compared to that of GR, and similar to that of GR/ZnO/MoS$_2$ trilayer. The results indicate that the enhanced interfacial interaction in 2D vdW trilayer heterostructure is irrelevant to the layer sequence.

We predict that the strengthened interaction between adjacent layers in GR/ZnO/MoS$_2$ trilayer by addition of other layer is general in 2D vdWs multilayer heterostructure. 2D AlN and g-C$_3$N$_4$ sheets, as two paradigms, are selected to substitute ZnO sheet for investigating the interfacial interaction in multilayer heterostructures. For this type of 2D vdWs multilayer heterostructure, the interfacial interaction strength can be visually reflected by the amount of charge transferred at the interfaces, i.e., the more charge transferred, the stronger coupling is. Figure S7 (Supporting Information) shows that the amount of charge transferred between adjacent layers in GR/AlN/MoS$_2$ trilayer is obviously larger than those in corresponding bilayers, indicating that the interfacial interaction in trilayer is stronger than in bilayer. This situation for the GR/g-C$_3$N$_4$/MoS$_2$ trilayer is more pronounced compared to the corresponding bilayers (Figure S8, Supporting Information). The greater charge transfer in vdWs trilayer indicates that the third layer put on bilayer will lead to stronger interlayer coupling between adjacent layers. These results provide convincing theoretical evidence that the interlayer coupling strength in 2D vdWs multilayer heterostructures can be effectively tuned by both changing layer number and choosing appropriate 2D materials.

In summary, we have demonstrated that the interlayer coupling in 2D few-layer vdWs heterostructures can be tuned by varying the layer number. Compared to that in bilayer, the interlayer coupling between neighboring layers in trilayer vdW heterostructures can be dramatically enhanced by stacking third layer, which is directly supported by short interlayer separations and more charge transfer. The enhanced vdWs interaction in trilayer profoundly alters its unique near-gap electronic structure, resulting into it having strong light absorption over a wide range (<700 nm). This indicates that 2D $MoS_2$-GR-based vdWs heterostructures are prime candidates for highly active photodetectors, high efficient solar energy harvesting and conversion. More importantly, the Dirac point of GR and band gaps of each layer and trilayer can be modulated by external electric field, making it very potential for optoelectronic devices. These findings suggest that $MoS_2$-GR-based vdWs heterostructures deserve concerted efforts to design and fabricate for applications in photodetector, solar energy harvesting and conversion, and photocatalysis.

## ASSOCIATED CONTENT

**Supporting Information**

Optimized geometric structures and interfacial charge redistributions of $ZnO/MoS_2$ bilayer vdW heterostructure in six different stacking patterns; band structures for pure and strained GR, ZnO and $MoS_2$ monolayers; DOS for $ZnO/MoS_2$ bilayer and $GR/ZnO/MoS2$ trilayer; projected band structures for $GR/ZnO/MoS_2$ trilayer under various vertical forward external fields; projected band structures for $GR/ZnO/MoS_2$ trilayer under various vertical reverse external fields; optimized geometric structures, three-dimensional charge density difference and planar averaged charge density difference for ZnO/GR, $GR/MoS_2$ and $ZnO/GR/MoS_2$ vdW heterostructures, DOS for $ZnO/GR/MoS_2$ trilayer and absorption spectra for monolayer GR and trilayer vdW heterostructures; optimized geometric structures, three-dimensional charge density difference and planar averaged charge density difference for GR/AlN, $GR/AlN/MoS_2$ and $AlN/MoS_2$ heterostructures; optimized geometric structures, three-dimensional charge density difference and planar averaged charge density difference for $GR/g-C_3N_4$, $GR/g-C_3N_4/MoS_2$ and $g-C_3N_4/MoS_2$ heterostructures. This material is available free of charge via the Internet at http://pubs.acs.org.

## AUTHOR INFORMATION

**Corresponding Authors**
*E-mail: wqhuang@hnu.edu.cn.
*E-mail: gfhuang@hnu.edu.cn.
**Notes**

The authors declare no competing financial interest.

**ACKNOWLEDGMENTS**

This work was supported by the National Natural Science Foundation of China (Grant Nos. 51471068 and 51525202).

**References**

**Uncategorized References**

(1) Geim, A.K.;Grigorieva, I.V. Nature, 2013, 499, 419-425.

(2) Novoselov, K.S.; Mishchenko, A.; Carvalho, A.;Neto, A.H.C. Science, 2016, 353, 461-+.

(3) Deng, D.H.; Novoselov, K.S.; Fu, Q.; Zheng, N.F.; Tian, Z.Q.;Bao, X.H. NAT. NANOTECHNOL, 2016, 11, 218-230.

(4) Li, G.; Luican, A.; Dos Santos, J.L.; Neto, A.C.; Reina, A.; Kong, J.;Andrei, E. NatPh, 2010, 6, 109-113.

(5) Yankowitz, M.; Xue, J.; Cormode, D.; Sanchez-Yamagishi, J.D.; Watanabe, K.; Taniguchi, T.; Jarillo-Herrero, P.; Jacquod, P.;LeRoy, B.J. NatPh, 2012, 8, 382-386.

(6) Kim, K.; Coh, S.; Tan, L.Z.; Regan, W.; Yuk, J.M.; Chatterjee, E.; Crommie, M.; Cohen, M.L.; Louie, S.G.;Zettl, A. Phys. Rev. Lett., 2012, 108, 246103.

(7) Havener, R.W.; Zhuang, H.; Brown, L.; Hennig, R.G.;Park, J. Nano Lett., 2012, 12, 3162-3167.

(8) Trambly de Laissardiere, G.; Mayou, D.;Magaud, L. Nano Lett., 2010, 10, 804-808.

(9) Luican, A.; Li, G.; Reina, A.; Kong, J.; Nair, R.; Novoselov, K.S.; Geim, A.K.;Andrei, E. Phys. Rev. Lett., 2011, 106, 126802.

(10) Novoselov, K.; McCann, E.; Morozov, S.; Fal'ko, V.I.; Katsnelson, M.; Zeitler, U.; Jiang, D.; Schedin, F.;Geim, A. NatPh, 2006, 2, 177-180.

(11) Bistritzer, R.;MacDonald, A.H. Proceedings of the National Academy of Sciences, 2011, 108, 12233-12237.

(12) Dean, C.; Wang, L.; Maher, P.; Forsythe, C.; Ghahari, F.; Gao, Y.; Katoch, J.; Ishigami, M.; Moon, P.;Koshino, M. Nature, 2013, 497, 598-602.

(13) Hunt, B.; Sanchez-Yamagishi, J.; Young, A.; Yankowitz, M.; LeRoy, B.J.; Watanabe, K.; Taniguchi, T.; Moon, P.; Koshino, M.;Jarillo-Herrero, P. Science, 2013, 340, 1427-1430.

(14) Ponomarenko, L.; Gorbachev, R.; Yu, G.; Elias, D.; Jalil, R.; Patel, A.; Mishchenko, A.; Mayorov, A.; Woods, C.;Wallbank, J. Nature, 2013, 497, 594-597.

(15) Lin, Y.-C.; Ghosh, R.K.; Addou, R.; Lu, N.; Eichfeld, S.M.; Zhu, H.; Li, M.-Y.; Peng, X.; Kim, M.J.; Li, L.-J.; Wallace, R.M.; Datta, S.;Robinson, J.A. Nature Communications, 2015, 6.

(16) Withers, F.; Del Pozo-Zamudio, O.; Mishchenko, A.; Rooney, A.P.; Gholinia, A.; Watanabe, K.; Taniguchi, T.; Haigh, S.J.; Geim, A.K.; Tartakovskii, A.I.;Novoselov, K.S. Nature Materials, 2015, 14, 301-306.

(17) He, J.; Kumar, N.; Bellus, M.Z.; Chiu, H.-Y.; He, D.; Wang, Y.;Zhao, H. Nature Communications, 2014, 5.

(18) Lee, C.-H.; Lee, G.-H.; van der Zande, A.M.; Chen, W.; Li, Y.; Han, M.; Cui, X.; Arefe, G.; Nuckolls, C.; Heinz, T.F.; Guo, J.; Hone, J.;Kim, P. NAT. NANOTECHNOL, 2014, 9, 676-681.

(19) Massicotte, M.; Schmidt, P.; Vialla, F.; Schaedler, K.G.; Reserbat-Plantey, A.; Watanabe, K.; Taniguchi, T.; Tielrooij, K.J.;Koppens, F.H.L. NAT. NANOTECHNOL, 2016, 11, 42-+.

(20) Dai, S.; Ma, Q.; Liu, M.K.; Andersen, T.; Fei, Z.; Goldflam, M.D.; Wagner, M.; Watanabe, K.; Taniguchi, T.; Thiemens, M.; Keilmann, F.; Janssen, G.C.A.M.; Zhu, S.E.; Jarillo-Herrero, P.; Fogler,

M.M.;Basov, D.N. NAT. NANOTECHNOL, 2015, 10, 682-686.
(21) Vu, Q.A.; Shin, Y.S.; Kim, Y.R.; Nguyen, V.L.; Kang, W.T.; Kim, H.; Luong, D.H.; Lee, I.M.; Lee, K.; Ko, D.-S.; Heo, J.; Park, S.; Lee, Y.H.;Yu, W.J. Nature communications, 2016, 7, 12725-12725.
(22) Costanzo, D.; Jo, S.; Berger, H.;Morpurgo, A.F. NAT. NANOTECHNOL, 2016, 11, 339-+.
(23) Fogler, M.M.; Butov, L.V.;Novoselov, K.S. Nature Communications, 2014, 5.
(24) Hong, X.; Kim, J.; Shi, S.-F.; Zhang, Y.; Jin, C.; Sun, Y.; Tongay, S.; Wu, J.; Zhang, Y.;Wang, F. NAT. NANOTECHNOL, 2014, 9, 682-686.
(25) Srivastava, A.; Sidler, M.; Allain, A.V.; Lembke, D.S.; Kis, A.;Imamoglu, A. NAT. NANOTECHNOL, 2015, 10, 491-496.
(26) Deng, D.; Novoselov, K.S.; Fu, Q.; Zheng, N.; Tian, Z.;Bao, X. NAT. NANOTECHNOL, 2016, 11, 218-230.
(27) Li, D.; Cheng, R.; Zhou, H.; Wang, C.; Yin, A.; Chen, Y.; Weiss, N.O.; Huang, Y.;Duan, X. Nature Communications, 2015, 6.
(28) Duan, X.; Wang, C.; Pan, A.; Yu, R.;Duan, X. Chem. Soc. Rev., 2015, 44, 8859-8876.
(29) Georgiou, T.; Jalil, R.; Belle, B.D.; Britnell, L.; Gorbachev, R.V.; Morozov, S.V.; Kim, Y.-J.; Gholinia, A.; Haigh, S.J.; Makarovsky, O.; Eaves, L.; Ponomarenko, L.A.; Geim, A.K.; Novoselov, K.S.;Mishchenko, A. NAT. NANOTECHNOL, 2013, 8, 100-103.
(30) Yan, K.; Peng, H.; Zhou, Y.; Li, H.;Liu, Z. Nano Lett., 2011, 11, 1106-1110.
(31) Zhang, C.; Zhao, S.; Jin, C.; Koh, A.L.; Zhou, Y.; Xu, W.; Li, Q.; Xiong, Q.; Peng, H.;Liu, Z. Nature Communications, 2015, 6.
(32) Doganov, R.A.; O'Farrell, E.C.T.; Koenig, S.P.; Yeo, Y.; Ziletti, A.; Carvalho, A.; Campbell, D.K.; Coker, D.F.; Watanabe, K.; Taniguchi, T.; Castro Neto, A.H.;Ozyilmaz, B. Nature Communications, 2015, 6.
(33) Blochl, P.E. Physical Review B (Condensed Matter), 1994, 50, 17953-17979.
(34) Kresse, G.;Furthmuller, J. Computational Materials Science, 1996, 6, 15-50.
(35) Grimme, S. J. Comput. Chem., 2006, 27, 1787-1799.
(36) Fang, H.; Battaglia, C.; Carraro, C.; Nemsak, S.; Ozdol, B.; Kang, J.S.; Bechtel, H.A.; Desai, S.B.; Kronast, F.; Unal, A.A.; Conti, G.; Conlon, C.; Palsson, G.K.; Martin, M.C.; Minor, A.M.; Fadley, C.S.; Yablonovitch, E.; Maboudian, R.;Javey, A. Proc. Natl. Acad. Sci. U.S.A., 2014, 111, 6198-6202.
(37) Lee, J.H.; Avsar, A.; Jung, J.; Tan, J.Y.; Watanabe, K.; Taniguchi, T.; Natarajan, S.; Eda, G.; Adam, S.; Neto, A.H.C.;Oezyilmaz, B. Nano Lett., 2015, 15, 319-325.
(38) Xu, L.; Huang, W.-Q.; Wang, L.-L.; Tian, Z.-A.; Hu, W.; Ma, Y.; Wang, X.; Pan, A.;Huang, G.-F. Chem. Mater., 2015, 27, 1612-1621.
(39) Giovannetti, G.; Khomyakov, P.A.; Brocks, G.; Karpan, V.M.; van den Brink, J.;Kelly, P.J. Phys. Rev. Lett., 2008, 101.
(40) Dahal, A.; Addou, R.; Coy-Diaz, H.; Lallo, J.;Batzill, M. Apl Materials, 2013, 1.
(41) Massicotte, M.; Schmidt, P.; Vialla, F.; Schadler, K.G.; Reserbat-Plantey, A.; Watanabe, K.; Taniguchi, T.; Tielrooij, K.J.;Koppens, F.H.L. NAT. NANOTECHNOL, 2016, 11, 42-+.
(42) Mak, K.F.;Shan, J. Nature Photonics, 2016, 10, 216-226.
(43) Bradley, A.J.; Ugeda, M.M.; da Jornada, F.H.; Qiu, D.Y.; Ruan, W.; Zhang, Y.; Wickenburg, S.; Riss, A.; Lu, J.; Mo, S.-K.; Hussain, Z.; Shen, Z.-X.; Louie, S.G.;Crommie, M.F. Nano Lett., 2015, 15, 2594-2599.
(44) Jariwala, D.; Davoyan, A.R.; Tagliabue, G.; Sherrott, M.C.; Wong, J.;Atwater, H.A. Nano Lett., 2016, 16, 5482-5487.
(45) Wallbank, J.R.; Ghazaryan, D.; Misra, A.; Cao, Y.; Tu, J.S.; Piot, B.A.; Potemski, M.; Pezzini, S.; Wiedmann, S.; Zeitler, U.; Lane, T.L.M.; Morozov, S.V.; Greenaway, M.T.; Eaves, L.; Geim, A.K.; Fal'ko, V.I.;

Novoselov, K.S.;Mishchenko, A. Science, 2016, 353, 575-579.
(46) Britnell, L.; Gorbachev, R.V.; Jalil, R.; Belle, B.D.; Schedin, F.; Mishchenko, A.; Georgiou, T.; Katsnelson, M.I.; Eaves, L.; Morozov, S.V.; Peres, N.M.R.; Leist, J.; Geim, A.K.; Novoselov, K.S.;Ponomarenko, L.A. Science, 2012, 335, 947-950.
(47) Huang, L.; Huo, N.; Li, Y.; Chen, H.; Yang, J.; Wei, Z.; Li, J.;Li, S.-S. The Journal of Physical Chemistry Letters, 2015, 6, 2483-2488.

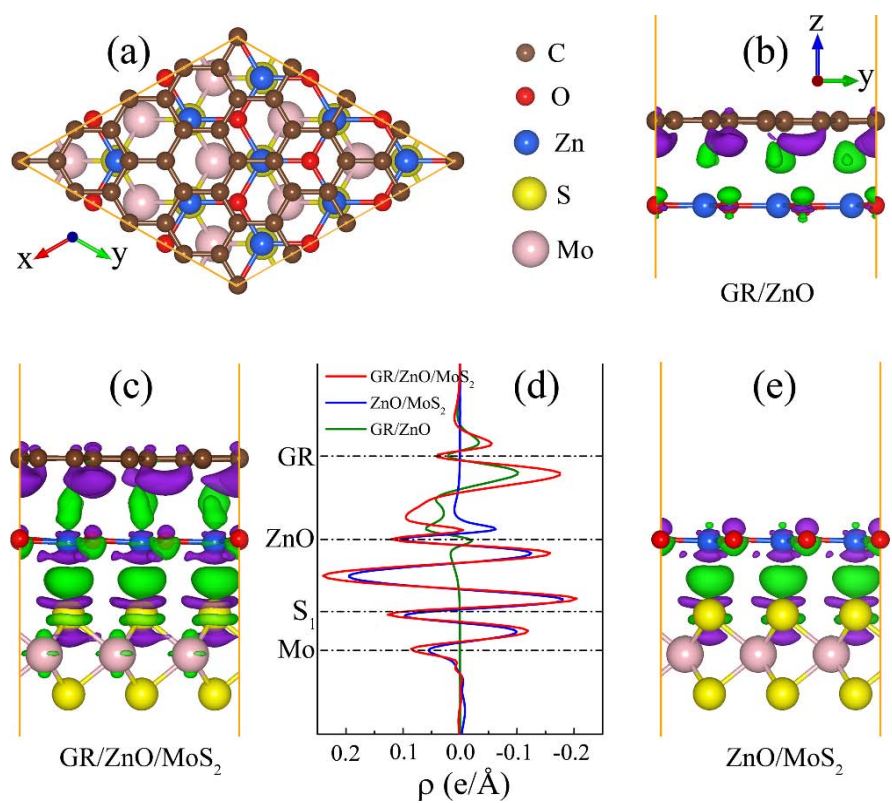

**Figure 1**. (a) Top view of 2D GR/ZnO/MoS$_2$ vdW heterostructure. Here, the small brown and red balls are C and O atoms; the middle blue balls are Zn atoms; and the large yellow and light magenta balls are Mo and S atoms, respectively. Three-dimensional charge density difference for (b) GR/ZnO, (c) GR/ZnO/MoS$_2$, and (e) ZnO/MoS$_2$ vdW heterostructures. The cyan and purple regions represent charge accumulation and depletion, respectively; the isosurface value is 0.003 $e$/Å3. Obviously, GR sheet strengthens the electron transfer between ZnO and MoS$_2$, especially the Mo atoms gain some electrons from S atoms at top layer. (d) The profile of the planar averaged charge density difference for the GR/ZnO/MoS$_2$, ZnO/MoS$_2$ and GR/ZnO vdW heterostructures as a function of position in the z-direction. The horizontal dashed-dot lines denote the central location of atomic layer of the GR, ZnO and MoS$_2$.

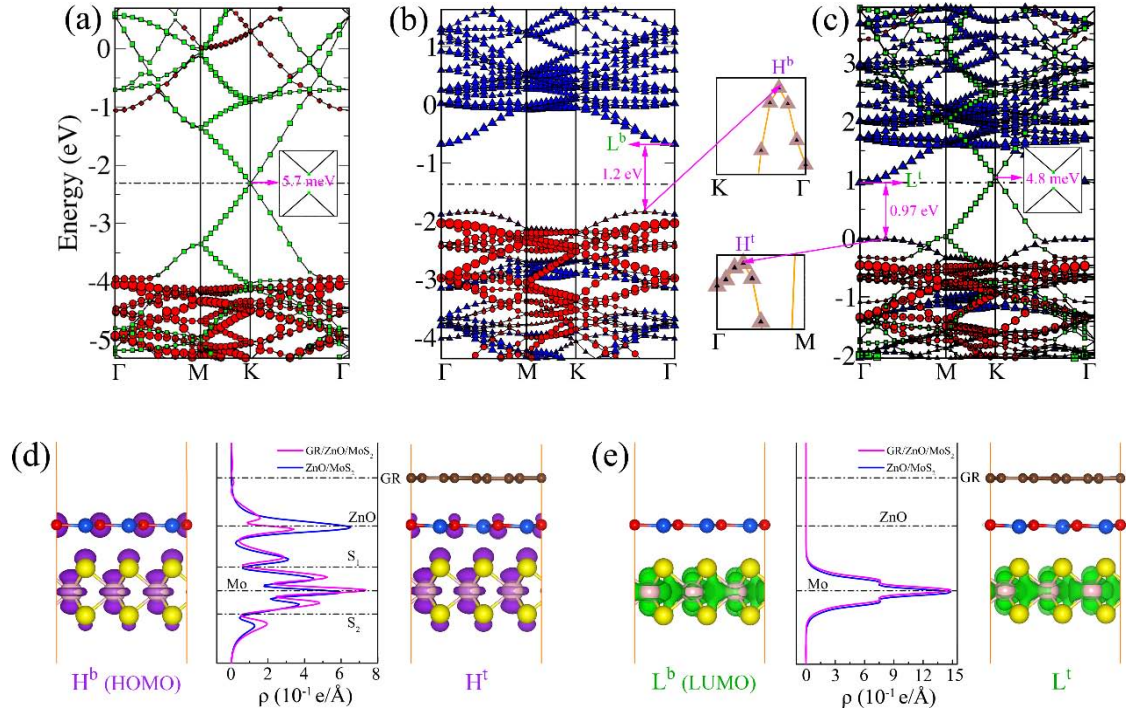

**Figure 2. Electronic properties of bilayer and trilayer vdW heterostructures**. (a-c) The projected band structures of GR/ZnO, ZnO/MoS$_2$ and GR/ZnO/MoS$_2$, which the bands dominated by GR, ZnO, and MoS$_2$ layers are plotted by green squares, red circles, and blue triangles, respectively. Figures between (b) and (c): Zoom-in the energy level including the VBM of ZnO/MoS$_2$ in bilayer (trilayer) vdW heterostructures. Relevant electronic parameters are also given in the Figure. The red dashed-dot lines are the Fermi level. Right (left) panel of (d,e), H$^b$ (H$^t$) and L$^b$ (L$^t$) and are the VBM and CBM of ZnO/MoS$_2$ in bilayer (trilayer) vdW heterostructures, respectively. Middle panel of (d,e), Profile of the planar averaged electronic density of H$^b$, H$^t$ (L$^b$, L$^t$) as a function of position in the z-direction. The horizontal dashed-dot lines are used to designate the central location of corresponding atomic layers in the trilayer vdW heterostructures, respectively.

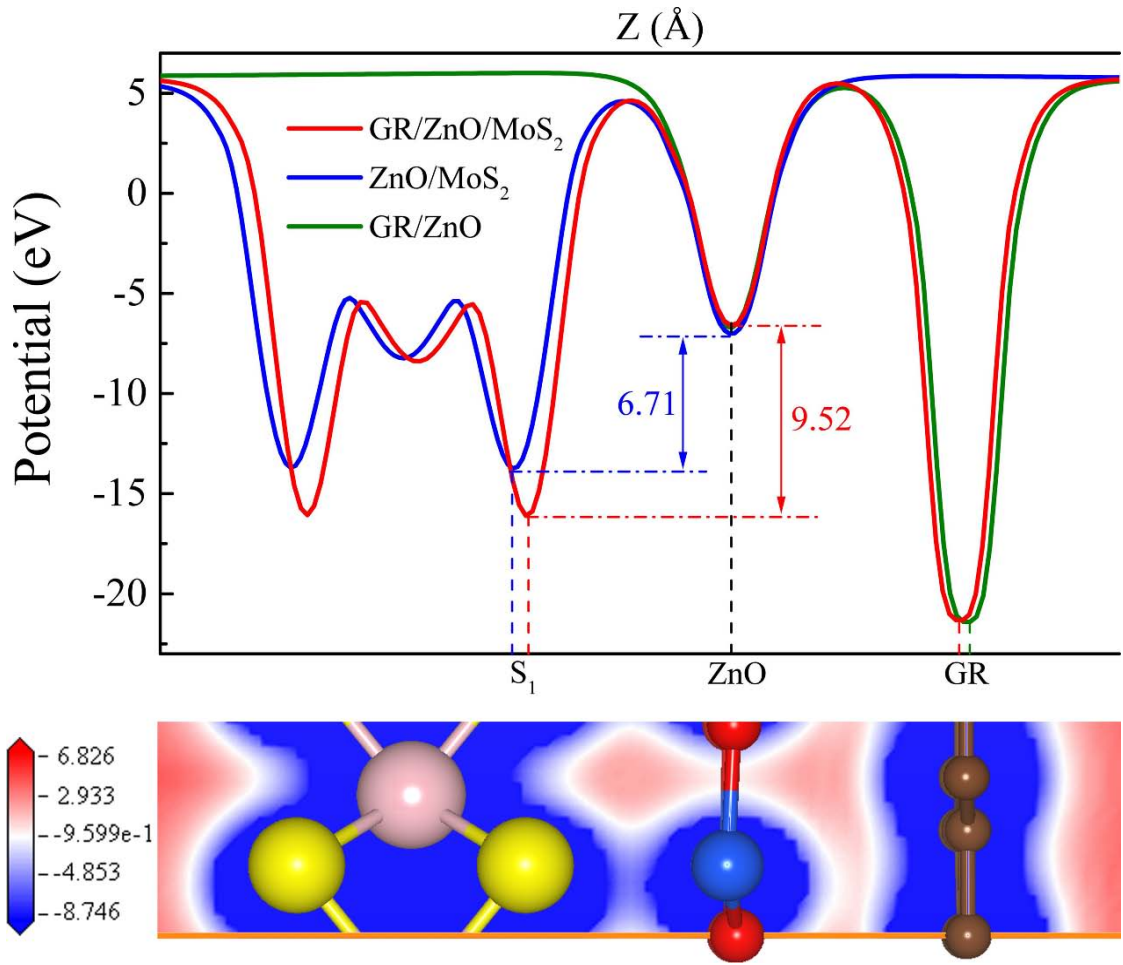

**Figure 3**. Upper panel: Profile of the planar averaged self-consistent electrostatic potential for the GR/ZnO, ZnO/MoS$_2$ and GR/ZnO/MoS$_2$ vdW heterostructures as a function of position in the z-direction. For comparison, the positions are aligned to the central location of 2D ZnO layer. As a visual guide, the change of interlayer separation in these vdW heterostructures is denoted by identifying the central location of S$_1$ and GR. Compared to bilayer ZnO/MoS$_2$ vdW heterostructure, the electrostatic potential at S atom in trilayer vdW heterostructure decreases obviously due to the stacking GR layer. Lower panel: Electrostatic potential of trilayer vdW heterostructure (one-third of the calculated supercell).

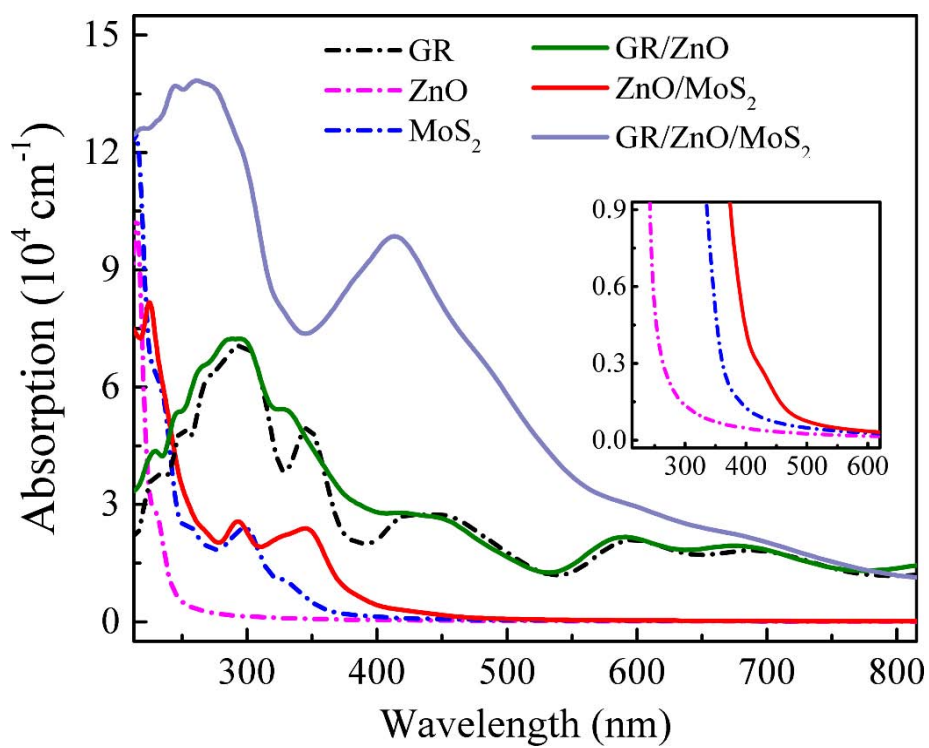

Figure 4. Absorption spectra of monolayer, bilayer and trilayer vdW heterostructures. Compared with the monolayers, the optical absorption property of bilayer vdW heterostructures only shows a slight improvement. Unexpected, the light absorption of trilayer vdW heterostructure is greatly enhanced in a wide range of visible light (300-750 nm), which have advantage in both visible light harvesting and conversion.

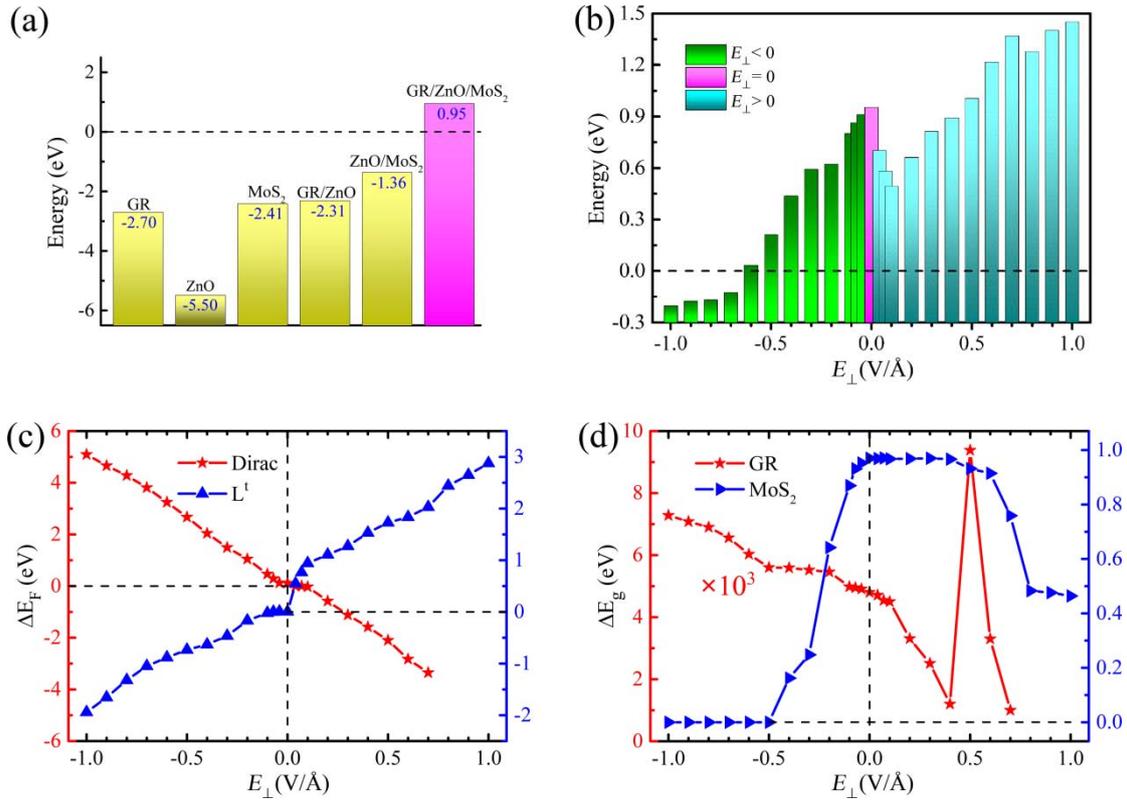

Figure 5. Electronic properties of trilayer vdW heterostructure as a function of the external field. **a**, Fermi energies of monolayers and vdW heterostructures without external field. For the GR/ZnO/MoS$_2$ trilayer, its Fermi energy is higher than zero (~0.95 eV). **b**, Fermi energy variation of GR/ZnO/MoS$_2$ trilayer under the external field. Here, E>0 denotes the direction of electric field from GR to MoS$_2$ layer, while E<0 presents reverse direction of electric field. **c**, Shift of Dirac and L$^t$ points with respect to Fermi level under the external field. **d**, Evolution of the band gap of GR and MoS$_2$ in the GR/ZnO/MoS$_2$ trilayer as a function of the external field.

Table 1: Interface binding energies ($E_{ad}$), mean distance ($d_1$, $d_2$) and Bader charge analysis of optimized GR/ZnO, ZnO/MoS$_2$ and GR/ZnO/MoS$_2$ composites.

| system | $E_{ad}$ (eV) | $d_1$ (Å) | $d_2$ (Å) | Bader charge (e) | | |
|---|---|---|---|---|---|---|
| | | | | GR | ZnO | MoS$_2$ |
| GR/ZnO | -0.902 | - | 3.153 | 0.027 | -0.027 | - |
| ZnO/MoS$_2$ | -4.234 | 2.890 | - | - | 0.066 | -0.066 |
| GR/ZnO/MoS$_2$ | -5.853 | 2.835 | 3.134 | 0.064 | 0.197 | -0.261 |

Table 2: Comparison of lattice constant (Å) and mismatch, band gap (eV) and work function (eV).

| system | lattice constant | mismatch | band gap | work function |
|---|---|---|---|---|
| GR | 2.467×4(9.868) | 1.1% | 0 | 4.51 |
| ZnO | 3.289×3(9.867) | 1.1% | 3.15 | 4.78 |
| MoS$_2$ | 3.190×3(9.570) | 2.0% | 1.97 | 5.52 |
| GR/ZnO | 9.762 | - | 0.12 | - |
| ZnO/MoS$_2$ | 9.762 | - | 1.48 | - |
| GR/ZnO/MoS$_2$ | 9.762 | - | 1.01($\Gamma$) | - |

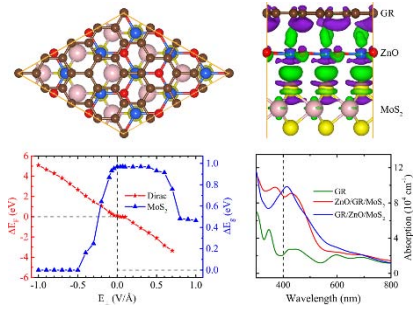

Table of Contents Graphic

# 2D MoS$_2$-Graphene-based multilayer van der Waals heterostructures: Enhanced charge transfer and optical absorption, and electric-field tunable Dirac point and band gap

Liang Xu[1,2], Wei-Qing Huang[1*], Wangyu Hu[2], Bing-Xin Zhou[1], Anlian Pan[1], Gui-Fang Huang[1#]

[1] *Department of Applied Physics, School of Physics and Electronics, Hunan University, Changsha 410082, China*

[2] *School of Materials Science and Engineering, Hunan University, Changsha 410082, China*

---

[*]. Corresponding author. *E-mail address:* wqhuang@hnu.edu.cn
[#]. Corresponding author. *E-mail address:* gfhuang@hnu.edu.cn

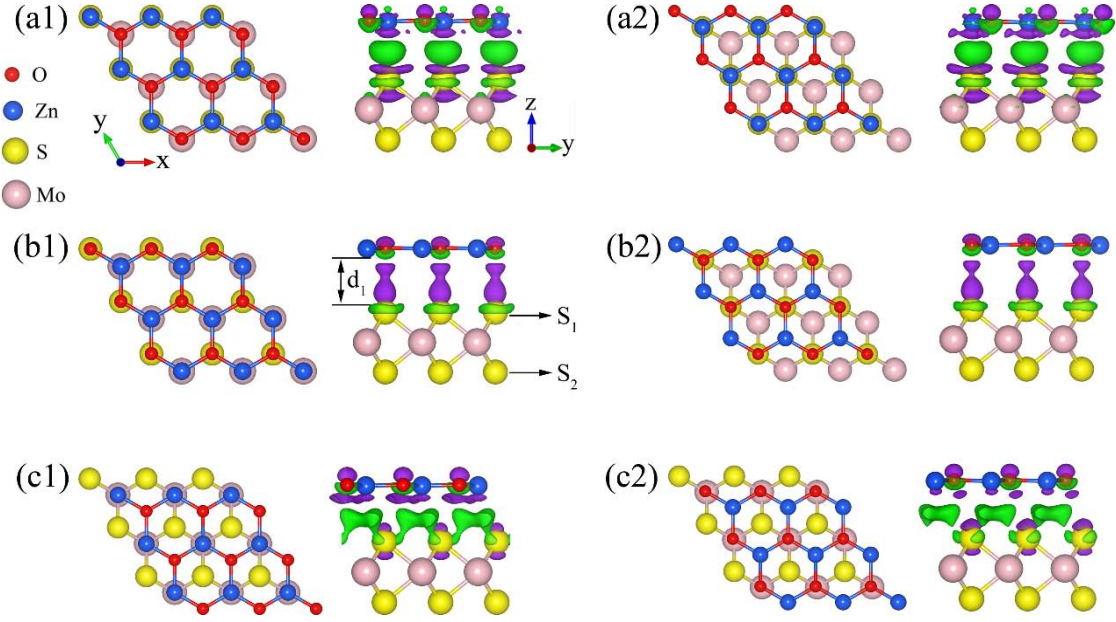

**Figure S1.** Top (left) and side (right) views of optimized structures of ZnO/MoS$_2$ bilayer vdW heterostructure in six stacking patterns. Clearly, different configurations induce different interfacial charge redistributions. The small red balls are O atoms, the middle blue balls are Zn atoms and the large yellow and light magenta balls are Mo and S atoms, respectively. The cyan and purple regions represent charge accumulation and depletion, respectively. The isosurface value is 0.003 e/Å$^3$.

**Table S1**: Energy difference $\Delta E$ (eV) between various configurations and the lowest energy configuration, the layer distance $d_1$ (Å) for ZnO/MoS$_2$ van der Waals heterostructures, as well as the Zn-O and Mo-S bond length for monolayers and heterostructures calculated by DFT-D3.

| system | configuration | $\Delta E$(eV) | $d_1$(Å) | $L_{Mo\text{-}S}$(Å) | $L_{Zn\text{-}O}$(Å) |
|---|---|---|---|---|---|
| ZnO | | | | | 1.894 |
| MoS$_2$ | | | | 2.413 | |
| ZnO/MoS$_2$ | a | 0.082 | 3.312 | 2.426 | 1.879 |
| | b | 0 | 2.890 | 2.422 | 1.879 |
| | c | 0.022 | 2.921 | 2.425 | 1.878 |
| | d | 0.006 | 2.910 | 2.423 | 1.880 |
| | e | 0.084 | 3.324 | 2.425 | 1.879 |
| | f | 0.021 | 2.917 | 2.424 | 1.878 |

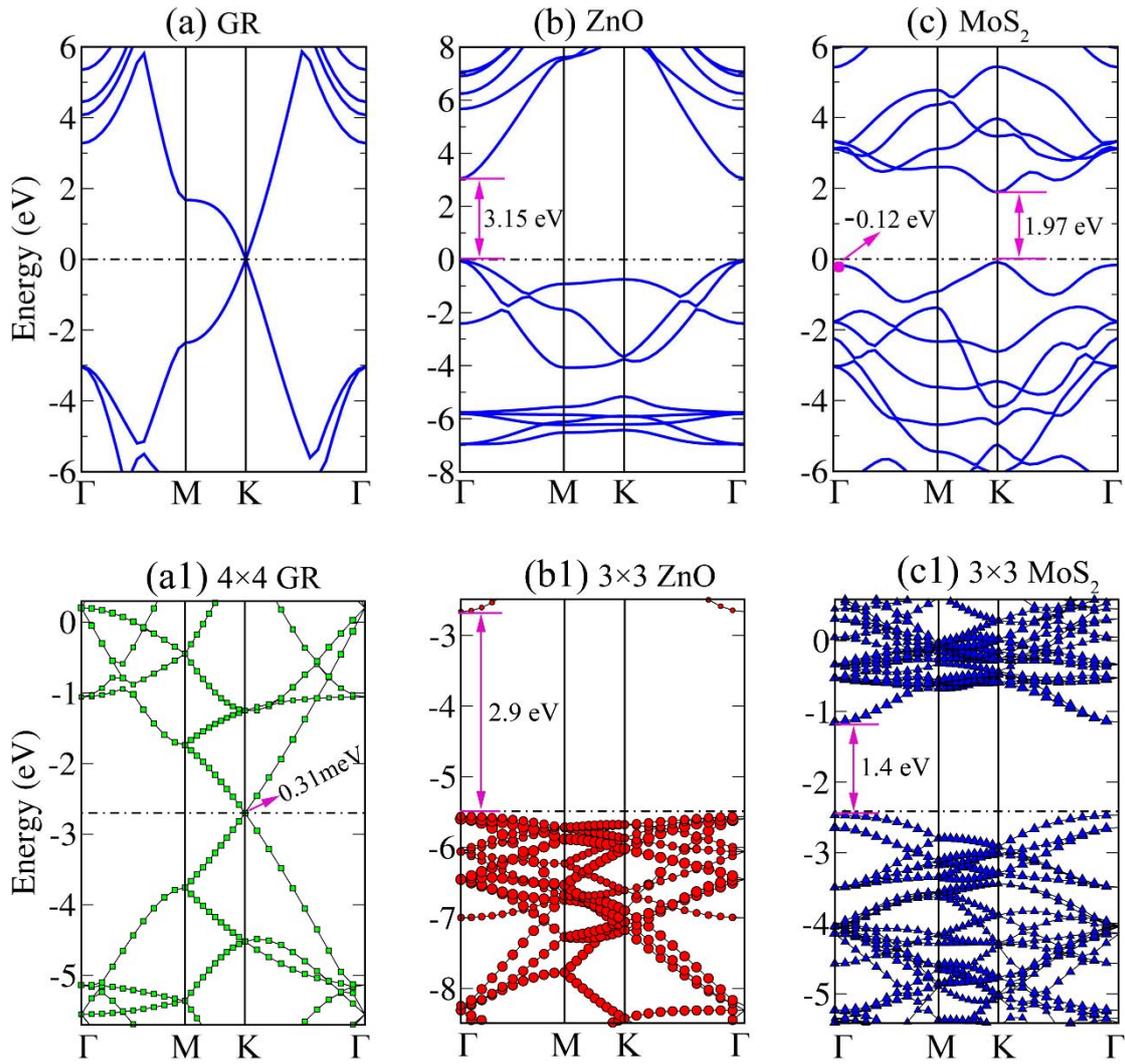

**Figure S2.** The band structures of pure GR (a), ZnO (b) and MoS$_2$ (c) monolayers. The GR sheet is a zero-gap semiconductor, ZnO and MoS$_2$ monolayers have the band gap of 3.15 eV and 1.97 eV, respectively, in well agreement with experimental and other theoretical values. The Fermi level is set at zero. The band structures of strained GR (4 × 4 unit cells) (a1), ZnO (3 × 3 unit cells) (b1) and MoS$_2$ (3 × 3 unit cells) (c1) having the same lattice constants with the corresponding layers in the GR/ZnO/MoS$_2$ trilayer vdW heterostructure. Due to the 1.1% strain, a very small band gap (about 0.31 meV) appears in the GR sheet, whereas the band gap of monolayer ZnO is decreased from 3.15 eV (without strain) to 2.9 eV. The relative large strain (2.0%) of monolayer MoS2 decreases its band gap from 1.97 to 1.4 eV. The black dashed-dot lines are the Fermi level.

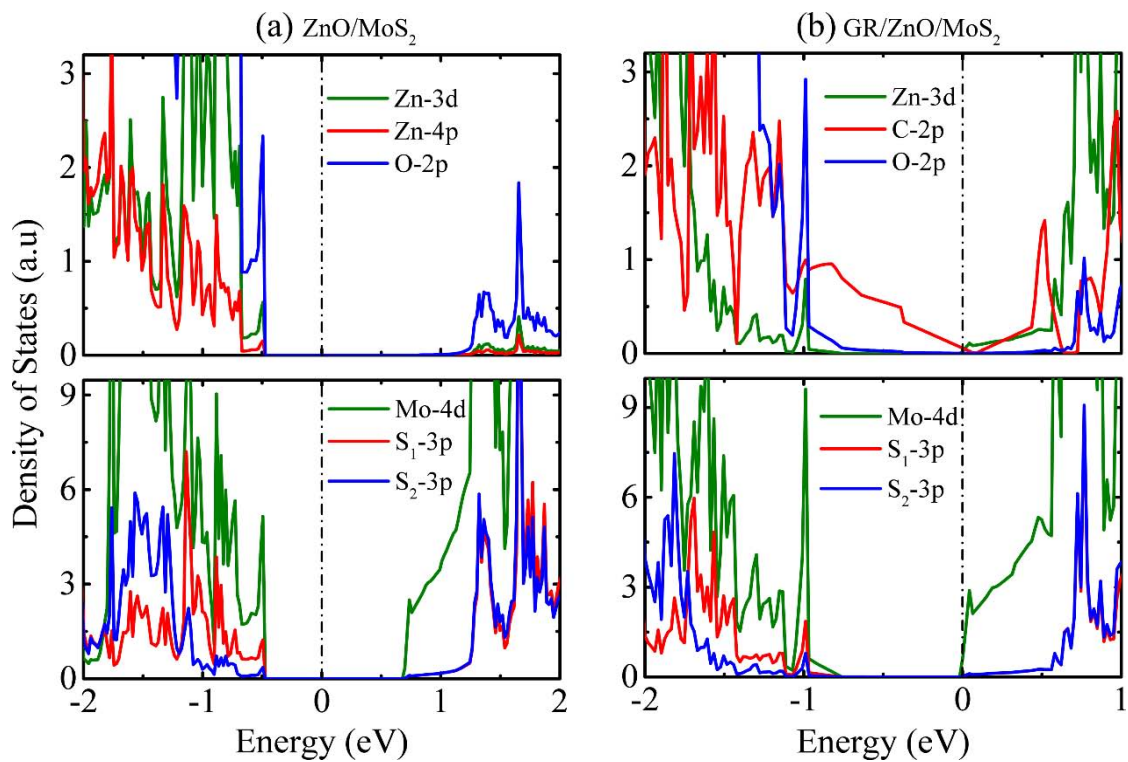

**Figure S3.** DOS for (a) ZnO/MoS$_2$ bilayer and (b) GR/ZnO/MoS2 trilayer. The vertical dotted-dash lines indicate the Fermi level.

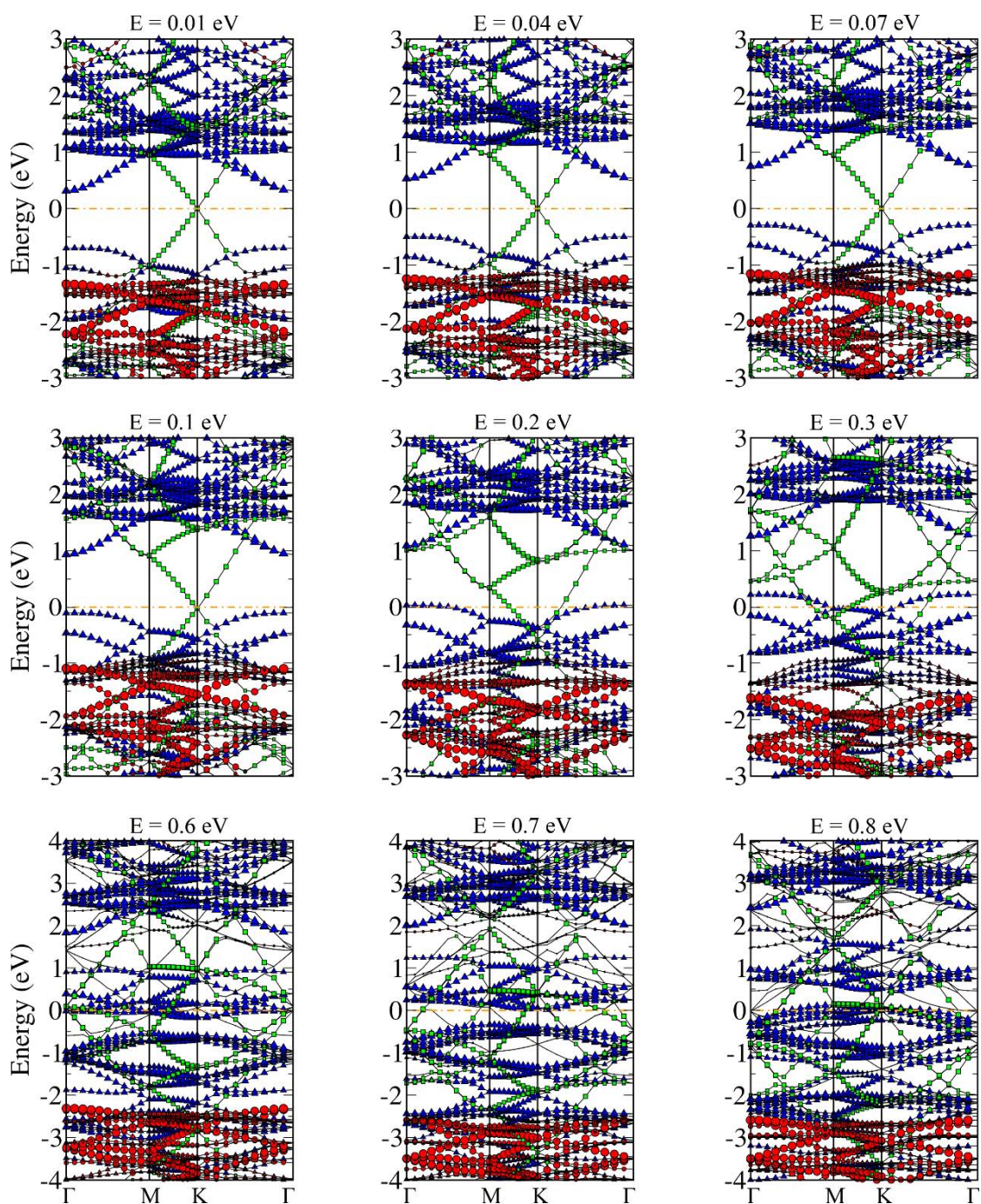

**Figure S4.** The projected band structures of GR/ZnO/MoS$_2$ trilayer under various vertical forward external fields. Here, the bands dominated by GR, ZnO, and MoS$_2$ layers are plotted by green squares, red circles and blue triangles, respectively. The Fermi level is set at zero.

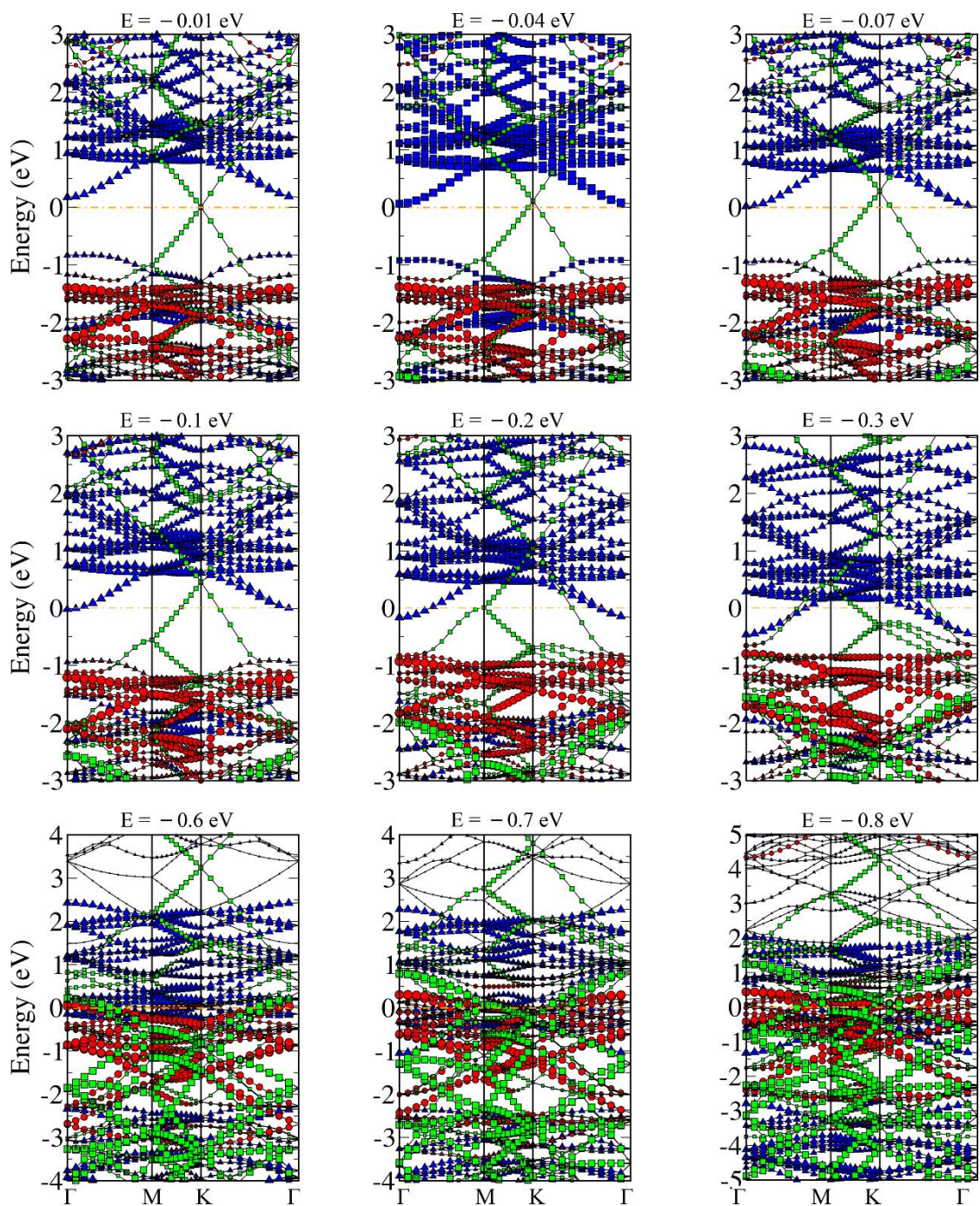

**Figure S5.** The projected band structures of GR/ZnO/MoS$_2$ trilayer under various vertical reverse external fields. Here, the bands dominated by GR, ZnO and MoS$_2$ layers are plotted by green squares, red circles and blue triangles, respectively. The Fermi level is set at zero.

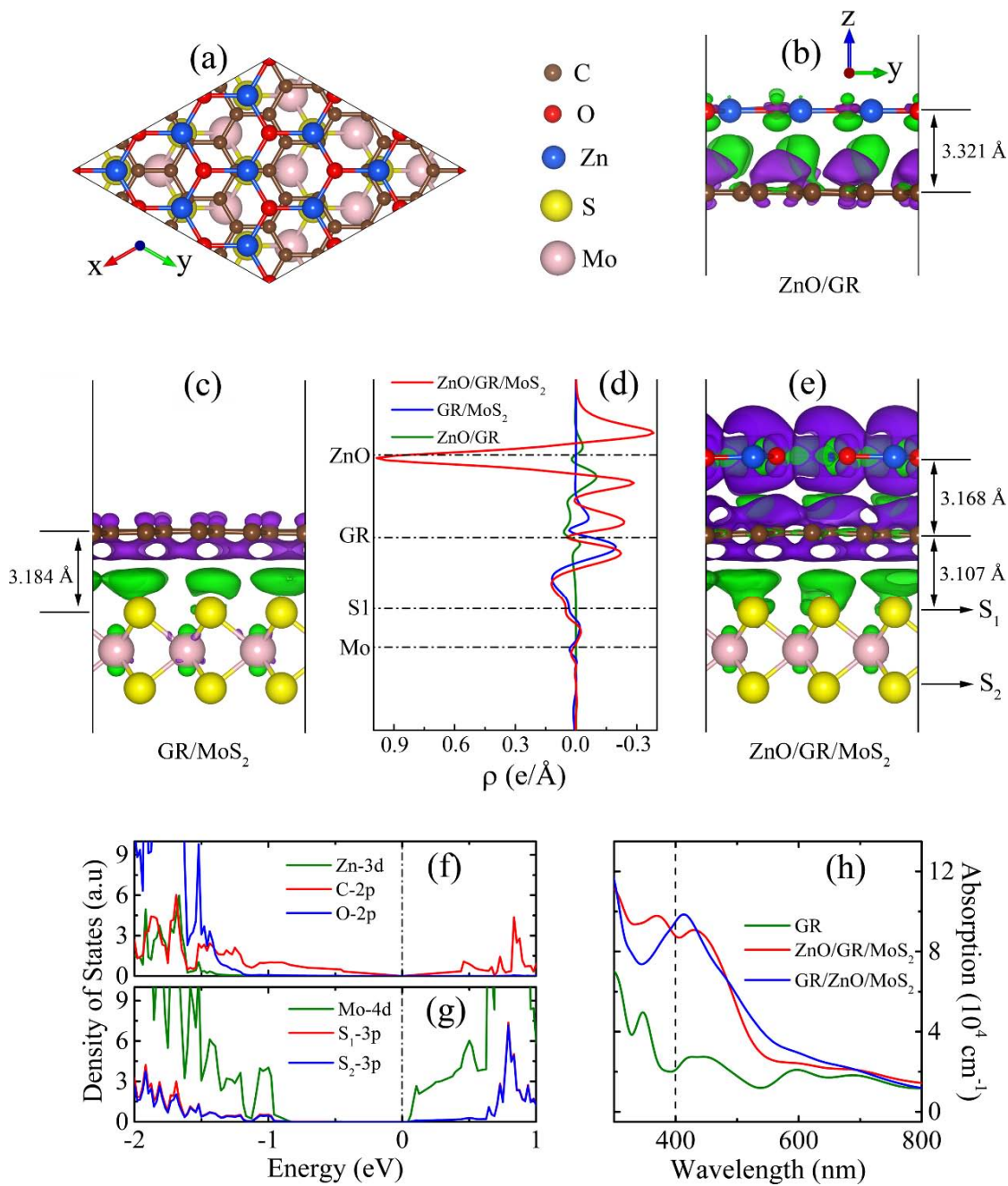

**Figure S6.** (a) Top views of optimized ZnO/GR/MoS$_2$ trilayer. Here, the small brown and red balls are C and O atoms, the middle blue balls are Zn atoms and the large yellow and light magenta balls are Mo and S atoms, respectively. Three-dimensional charge density difference for (b) ZnO/GR, (c) GR/MoS$_2$ and (e) ZnO/GR/MoS$_2$ vdW heterostructures. The isosurface value of 0.0002 e/Å$^3$. The cyan and purple regions represent charge accumulation and depletion, respectively. (d) The profile of the planar averaged charge density difference for the GR/ZnO/MoS2, ZnO/MoS2 and GR/ZnO vdW heterostructures as a function of position in the z-direction. (f, g) DOS for ZnO/GR/MoS$_2$ trilayer. The vertical dotted-dash lines indicate the Fermi level. (h) Absorption spectra of monolayer GR and trilayer vdW heterostructures.

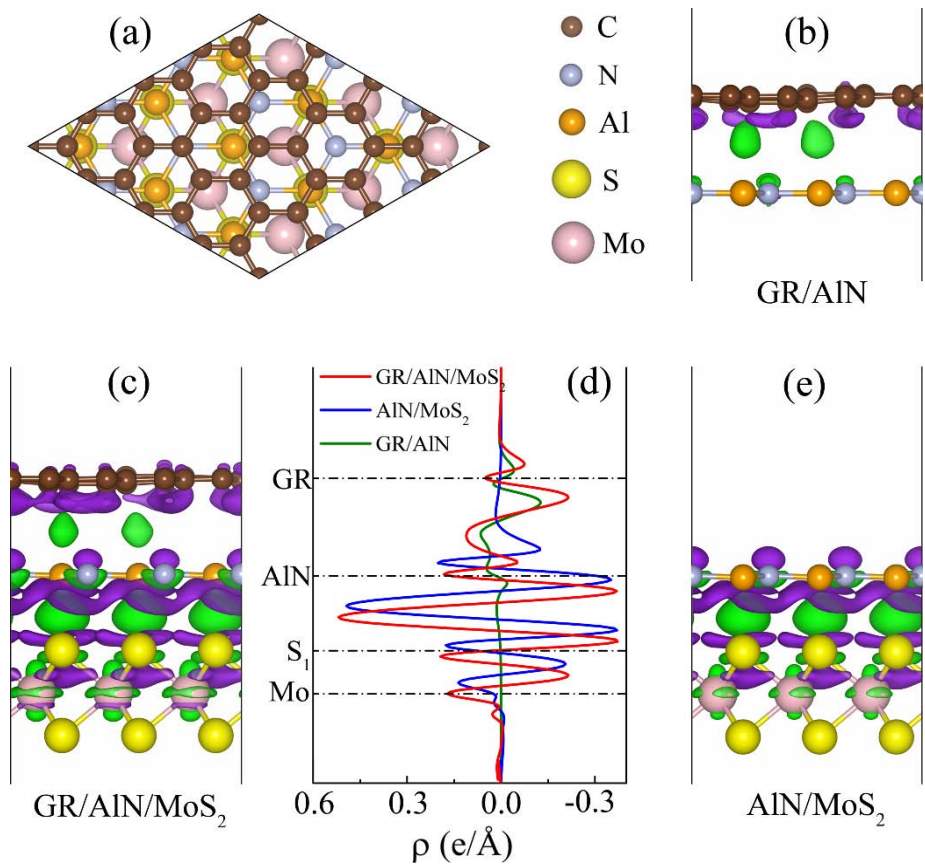

**Figure S7.** (a) Top views of optimized GR/AlN/MoS$_2$ trilayer. Here, the small brown and gray balls are C and N atoms, the middle orange balls are Al atoms and the large yellow and light magenta balls are Mo and S atoms, respectively. Three-dimensional charge density difference for (b) GR/AlN, (c) GR/AlN/MoS$_2$ and (e) AlN/MoS$_2$ heterostructures. The cyan and purple regions represent charge accumulation and depletion, respectively. The isosurface value is 0.003 e/Å$^3$. (d) The profile of the planar averaged charge density difference for the GR/AlN/MoS$_2$, AlN/MoS$_2$ and GR/AlN heterostructures as a function of position in the z-direction. The horizontal dash dot lines denote the central location of atomic layer of the GR, AlN and MoS2.

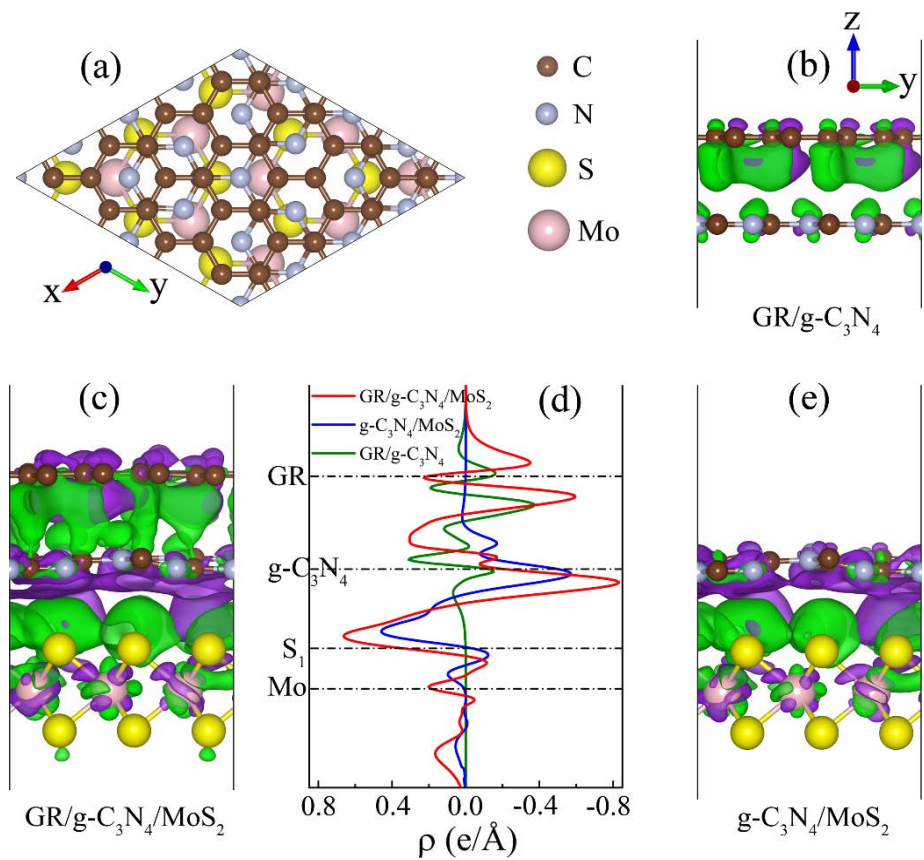

**Figure S8.** (a) Top views of optimized GR/g-$C_3N_4$/$MoS_2$ trilayer. Here, the small brown and gray balls are C and N atoms; and the large yellow and light magenta balls are Mo and S atoms, respectively. Three-dimensional charge density difference for (b) GR/g-$C_3N_4$, (c) GR/g-$C_3N_4$/$MoS_2$ and (e) g-$C_3N_4$/$MoS_2$ heterostructures. The cyan and purple regions represent charge accumulation and depletion, respectively. The isosurface value is 0.003 e/Å$^3$. (d) The profile of the planar averaged charge density difference for the GR/g-$C_3N_4$/$MoS_2$, g-$C_3N_4$/$MoS_2$ and GR/g-$C_3N_4$ heterostructures as a function of position in the z-direction. The horizontal dash dot lines denote the central location of atomic layer of the GR, g-$C_3N_4$ and MoS2.